\documentclass[letterpaper, 10 pt, conference]{ieeeconf}  

\IEEEoverridecommandlockouts                              
\usepackage{graphicx,amssymb,amstext}
\usepackage{amsmath}
\usepackage{tikz,lscape}
\usepackage{hyperref}
\usepackage{epsfig}
\usetikzlibrary{shapes,arrows}
\usepackage{verbatim}

\title{\LARGE \bf
Adaptive Continuous Homodyne Phase Estimation\\ Using Robust Fixed-Interval Smoothing
}

\author{Shibdas Roy$^{1}$*, Ian R. Petersen$^{2}$ and Elanor H. Huntington$^{3}$%
\thanks{$^{1}$S. Roy, $^{2}$I. R. Petersen and $^{3}$E. H. Huntington are with the School of Engineering and Information Technology, University of New South Wales, Canberra.}%
\thanks{*\tt\small shibdas.roy at student.adfa.edu.au}%
}

\begin{document}

\tikzstyle{block} = [draw, fill=white, rectangle, 
    minimum height=1em, minimum width=2em]
\tikzstyle{divide} = [draw, fill=white, circle, inner sep=0.5mm]
\tikzstyle{open}=[inner sep=1mm]
\tikzstyle{none}=[inner sep=0mm]

\maketitle
\thispagestyle{empty}
\pagestyle{empty}


\begin{abstract}

Adaptive homodyne estimation of a continuously evolving optical phase using time-symmetric quantum smoothing has been demonstrated experimentally to provide superior accuracy in the phase estimate compared to adaptive or non-adaptive estimation using filtering alone. Here, we illustrate how the mean-square error in the adaptive phase estimate may be further reduced below the standard quantum limit for the stochastic noise process considered by using a Rauch-Tung-Striebel smoother as the estimator, alongwith an optimal Kalman filter in the feedback loop. Further, the estimation using smoothing can be made robust to uncertainties in the underlying parameters of the noise process modulating the system phase to be estimated. This has been done using a robust fixed-interval smoother designed for uncertain systems satisfying a certain integral quadratic constraint.

\end{abstract}


\section{INTRODUCTION}

\bstctlcite{BSTcontrol}

Quantum parameter estimation (QPE) \cite{WM} involves estimating an unknown classical parameter of a quantum system and plays important role in various fields such as quantum computation \cite{HWA}, quantum key distribution \cite{IWY} and gravitational wave interferometry \cite{GMM}. A common and technologically relevant example of QPE is estimating an optical phase. The fundamental limit to the precision of the phase estimate is set by Heisenberg's uncertainty principle \cite{GLM}. On the other hand, the standard quantum limit (SQL) is the minimum level of quantum noise that can be obtained using standard approaches not involving real-time feedback.

Since the phase of an electromagnetic field cannot be measured directly, all phase-measurement schemes employ measurement of some other quantity, that necessarily introduces uncertainty in the phase estimate. The standard method of measuring the phase of a signal is the \emph{heterodyne} scheme, where the signal is combined with a strong local-oscillator (LO) field detuned from the signal, resulting in an introduced excess uncertainty scaling as $1/\mathcal{N}$ ($\mathcal{N} := |\alpha|^2$ is the mean photon number). By contrast, \emph{homodyne} scheme introduces greatly reduced uncertainty, in cases where there is some a-priori knowledge about the phase, by using an LO phase that is $\pi/2$ out of phase with the signal. Moreover, by \emph{adapting} the LO phase using feedback during the measurement, it is possible to further reduce the excess uncertainty to obtain a mean-square estimation error lower than the SQL \cite{HMW,WK1,WK2,MA} and even attain the theoretical limit \cite{BW1}.

However, these works were based on \emph{single-shot} measurements of \emph{fixed} unknown phase. Practically, it is more relevant to be able to keep track of a \emph{time-varying} phase instead \cite{BW2,TSL}. There are a number of ways of estimating a classical process dynamically coupled to a quantum system under continuous measurement, viz. prediction or filtering, smoothing and retrodiction \cite{MT}. Smoothing, in particular, is an estimation technique, that uses both past and future measurements, and, therefore, yields a more accurate estimate as compared to only filtering, that uses only past measurements. However, it is essentially a non-causal method that cannot be used in real-time but is used for offline data processing or with a delay with respect to the estimation time.

The \emph{fixed-interval} smoothing problem \cite{LK,JSM,WWS} involves measurements over a given fixed time-interval $T$. One solution to the fixed-interval smoothing problem is the Mayne-Fraser two-filter smoother \cite{DQM,DCF,RKM}, that uses, in addition to a forward-time Kalman filter, a backward-time Kalman filter, also known as an ``information filter" \cite{FP} and finally combines the two estimates to yield the optimal smoothed estimate. Rauch, Tung and Striebel combined the information filter and the smoother into a single backward smoother \cite{RTS,LXP}.

The first experimental demonstration of adaptive quantum phase estimation of a continuously varying phase using quantum smoothing was presented in Ref. \cite{TW}, where an estimate could be obtained with a mean-square error of up to $2.24 \pm 0.14$ times smaller than the SQL. The experiment used a classical stochastic Ornstein-Uhlenbeck (OU) noise process to modulate the signal phase to be estimated. The authors have previously shown in Ref. \cite{RPH} that the feedback filter used in Ref. \cite{TW} is only optimal when the noise process is a Wiener process and the measurement is assumed to be linear and that using an optimal Kalman filter, instead, significantly reduces the mean-square error in the phase estimate. Here, we show that using a Rauch-Tung-Striebel (RTS) smoother, in addition to the Kalman filter in the feedback loop, improves the accuracy of the phase estimate as compared to the (offline) estimator used in Ref. \cite{TW}.

It is desirable to make the estimation process robust to uncertainties
in the underlying parameters of the noise process, since it is physically unreasonable to specify these parameters accurately. Significant effort has been put on developing robust approaches to QPE \cite{JS,PWL}. The authors have illustrated in Ref. \cite{RPH} how the feedback filter and, therefore, the precision of the phase estimate can be made robust to uncertainty in one of the underlying parameters, based on a  guaranteed cost robust filtering approach \cite{PM}. Here, the optimal RTS smoother is made robust to parameter uncertainty by applying the fixed-interval robust smoothing theory described in Ref. \cite{MSP} for continuous-time uncertain systems satisfying a certain integral quadratic constraint.

\section{STANDARD QUANTUM LIMIT FOR ORNSTEIN-UHLENBECK PROCESS}

The standard quantum limit plays an important role as a benchmark for the quality of a measurement and is set by the minimum error in phase estimation that can be obtained using perfect heterodyne technique, or in other words, a non-adaptive filtering scheme. The case when the signal phase varies as a Weiner process has been considered in Ref. \cite{BW2}, where the minimum variance was obtained to be $\sqrt{\kappa}/(\sqrt{2}|\alpha|)$. In this section, we shall consider the case when the signal phase varies as an OU process as in Ref. \cite{TW}. We deduce the minimum error covariance for the case of OU noise using the standard optimal filtering approach rather than the method used in Ref. \cite{BW2}.

The OU noise process under consideration is \cite{TW,RPH}:
\begin{equation}\label{eq:ou_noise}
\dot{\phi}(t) = -\lambda\phi(t) + \sqrt{\kappa}v(t),
\end{equation}
where $\phi(t)$ is the system phase to be estimated, $\lambda > 0$ is the mean reversion rate, $\kappa > 0$ is the inverse coherence time and $v(t)$ is a zero-mean white Gaussian noise with unity amplitude.

\begin{figure}[!b]
\centering
\begin{tikzpicture}[scale=0.6]
	\node [open] (signal) at (1,0) {\footnotesize Signal};
    \node [block] (eom) at (3,0) {\footnotesize EOM};
	\node [block,align=center] (ou) at (3,2) {\footnotesize OU noise};
    \node [none] (bs) at (5,0) {$\diagup$};
    \node [open] (vacuum) at (5,-1.5) {\footnotesize $\delta v$};
    \node [block] (hd2) at (7,0) {\footnotesize HD2};
    \node [none] (mirror) at (5,2) {$\diagup$};
    \node [block] (hd1) at (7,2) {\footnotesize HD1};
    \node [divide] (divide) at (9,1) {\footnotesize $\div$};
    \node [block] (arctan) at (11,1) {\footnotesize arctan};
    \node [block] (lpf) at (13,1) {\footnotesize LPF};
    \node [open] (output) at (14.5,1) {\footnotesize $\hat{\phi}$};
    
    \draw [->] (signal) -- (eom);
    \draw [->] (ou) -- node [left, near end] {\footnotesize $e^{i\phi}$} (eom);
    \draw [->] (eom) -- node [above] {\footnotesize $|\alpha|e^{i\phi}$} (bs);
    \draw [->] (bs) -- (hd2); 
    \draw [->] (hd2) -| node [left, near end] {\footnotesize $I_2$} (divide);
    \draw [->,dashed] (vacuum) -- (bs);
    \draw [->] (bs) -- (mirror);
    \draw [->] (mirror) -- (hd1);
    \draw [->] (hd1) -| node [left, near end] {\footnotesize $I_1$} (divide);
    \draw [->] (divide) -- (arctan);
    \draw [->] (arctan) -- node [above] {\footnotesize $\vartheta$} (lpf);
    \draw [->] (lpf) -- (output);
\end{tikzpicture}
\caption{Block diagram of the dual-homodyne scheme for deducing the SQL for OU noise.}
\label{fig:dual_hd_sql}
\end{figure}
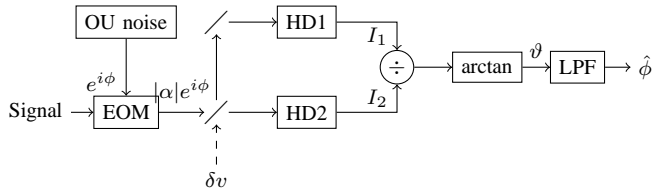

In our analysis here, we use the fact that the heterodyne scheme of measurement is, in principle, equivalent to, and incurs the same noise penalty as, \emph{dual-homodyne} scheme \cite{TW}, such as the schematic depicted in Fig. \ref{fig:dual_hd_sql}. We model the OU process as a signal at the input being phase-modulated using an electro-optic modulator (EOM) that is driven by an OU noise source. The modulated signal is then split using a $50-50$ beamsplitter into two arms each with a homodyne detector (HD1 and HD2, respectively, with the LO phase of HD1 $\pi/2$ out of phase with that of HD2). The ratio of the output signals of the two arms goes to an \emph{arctan} block, the output of which is fed into a low-pass filter (LPF). The filter for the case in Ref. \cite{TW} is sub-optimal, and we use here a Kalman filter, instead, to obtain the minimum error-covariance that determines the SQL for the OU process.

The output signals of the two arms are:
\begin{eqnarray*}
I_1 = \frac{1}{\sqrt{2}} \left( 2|\alpha| \sin\phi + n_1 + n_2 \right), \\ 
I_2 = \frac{1}{\sqrt{2}} \left( 2|\alpha| \cos\phi + n_3 - n_4 \right),
\end{eqnarray*}
where $n_1$ and $n_3$ are measurement noises of the two homodyne detectors, respectively, and $n_2$ and $n_4$ are the noises arising from the vacuum entering the empty port of the input beamsplitter corresponding to the two arms, respectively. All these noises are assumed to be zero-mean white Gaussian.

The output of the arctan block is:
\begin{equation}
\vartheta = \arctan \left( \frac{2|\alpha|\sin\phi + n_1 + n_2}{2|\alpha|\cos\phi + n_3 - n_4} \right).
\end{equation}

A Taylor series expansion up to first-order terms of the right-hand side yields:
\begin{equation}\label{eq:dual_hd}
\vartheta \approx \phi + \frac{1}{2|\alpha|}n_1 + \frac{1}{2|\alpha|}n_2.
\end{equation}

The transfer function of the low-pass filter for the case of Ref. \cite{TW} is \cite{RPH}:
\begin{equation}\label{eq:prl_filter}
G(s) = \frac{\hat{\phi}}{\vartheta} = \frac{\chi}{s+\chi}.
\end{equation}

We can determine the error covariance when that filter is used as follows. The system augmented with the filter may be represented by the state-space model:
\begin{equation}
\mathbf{\dot{\overline{x}}} = \mathbf{\overline{A}\, \overline{x}} + \mathbf{\overline{B}\, \overline{w}},
\end{equation}
where

\qquad \qquad \( \mathbf{\overline{x}} = 
\left[ \begin{array}{c}
\phi \\
\hat{\phi}
\end{array} \right] \)
\qquad and \qquad
\( \mathbf{\overline{w}} = 
\left[ \begin{array}{c}
v \\
n_1 \\ 
n_2 \\ 
n_3 \\ 
n_4
\end{array} \right]. \)

\vspace*{2mm}

From (\ref{eq:ou_noise}), (\ref{eq:dual_hd}) and (\ref{eq:prl_filter}), we get:

\begin{align}
\textsf{Process:}& \quad \dot{\phi} = -\lambda\phi + \sqrt{\kappa}v, \\
\textsf{Filter:}& \quad \dot{\hat{\phi}} = - \chi\hat{\phi} + \chi\phi + \frac{\chi}{2|\alpha|}n_1 + \frac{\chi}{2|\alpha|}n_2.
\end{align}

Thus, we have:
\small

\( \mathbf{\overline{A}} = 
\left[ \begin{array}{cc}
-\lambda & 0 \\
\chi & -\chi
\end{array} \right] \)
 and 
\( \mathbf{\overline{B}} = 
\left[ \begin{array}{ccccc}
\sqrt{\kappa} & 0 & 0 & 0 & 0\\
0 & \frac{\chi}{2|\alpha|} & \frac{\chi}{2|\alpha|} & 0 & 0
\end{array} \right]. \)

\normalsize
\vspace*{2mm}

The steady-state state covariance matrix $\mathbf{P}_S$ is obtained by solving the \emph{Lyapunov equation}:
\begin{equation}\label{eq:lyapunov}
\mathbf{\overline{A}P}_S + \mathbf{P}_S\mathbf{\overline{A}}^T + \mathbf{\overline{B}\, \overline{B}}^T = 0,
\end{equation}
where $\mathbf{P}_S$ is the symmetric matrix
\[ 
\mathbf{P}_S = E(\mathbf{\overline{x}\, \overline{x}}^T) =
\left[ \begin{array}{cc}
P_1 & P_2 \\
P_2 & P_3
\end{array} \right].
\]

Upon solving (\ref{eq:lyapunov}), we get
\begin{align*}
P_1& = \frac{\kappa}{2\lambda},\\
P_2& = \frac{\chi\kappa}{2\lambda(\lambda + \chi)},\\
P_3& = \frac{\chi}{2}\left\lbrace \frac{\kappa}{\lambda (\lambda + \chi)} + \frac{1}{2|\alpha|^2}\right\rbrace.
\end{align*}

The estimation error can be written as:
\[ \mathbf{e} = \phi - \hat{\phi} = [1 \, -1]\mathbf{\overline{x}}, \]
which is mean zero since all of the quantities determining $\mathbf{e}$ are mean zero.

The error covariance is then given as:

\begin{align*} 
\sigma^2& = E(\mathbf{ee}^T) = [1 \, -1]E(\mathbf{\overline{x}\, \overline{x}}^T)
\left[ \begin{array}{c}
1 \\
-1
\end{array} \right] \\ 
& =
[1 \, -1]
\left[ \begin{array}{cc}
P_1 & P_2 \\
P_2 & P_3
\end{array} \right]
\left[ \begin{array}{c}
1 \\
-1
\end{array} \right]
= P_1 - 2P_2 + P_3.
\end{align*}

Thus, we obtain:
\begin{equation}\label{eq:ou_filter_prl}
\sigma^2 = \frac{\kappa}{2(\lambda + \chi)} + \frac{\chi}{4|\alpha|^2}.
\end{equation}

Note that when $\lambda = 0$, the above expression for the error reduces to that given by Eq. (3.8) from Ref. \cite{BW2}.

By contrast, the optimal filter is given by the Kalman filter, which may be determined in steady-state from the algebraic Riccati equation, which for the process given by (\ref{eq:ou_noise}) and measurement given by (\ref{eq:dual_hd}) is:
\begin{equation}
-2\lambda P - 2|\alpha|^2 P^2 + \kappa = 0.
\end{equation}

The stabilising solution of the above equation for $P$ is:
\begin{equation}\label{eq:ou_filter_kalman}
\boxed{P = \frac{-\lambda + \sqrt{\lambda^2 + 2\kappa|\alpha|^2}}{2|\alpha|^2}.}
\end{equation}

This, being the minimum error that can be obtained without feedback, determines the standard quantum limit for OU noise process. Note that when $\lambda = 0$, we get $P = \sqrt{\kappa}/(\sqrt{2}|\alpha|)$, as expected.

The error-covariance given by (\ref{eq:ou_filter_prl}) is obtained when using the filter from Ref. \cite{TW}, that is only optimal when assuming Wiener noise but not so for the more general OU noise, since it has only one variable $\chi$ that controls both the gain and the corner frequency of the filter. On the contrary, in deriving the error covariance given by (\ref{eq:ou_filter_kalman}) of the Kalman filter, no such assumptions were made, so that the gain and the corner frequency of the filter were allowed to be two independent values, thereby yielding lower mean-square error in the estimate than in the former case.

\section{ADAPTIVE PHASE ESTIMATION USING SMOOTHER FROM REF. \cite{TW}}\label{sec:ape_prl}

In this section, we consider the (offline) estimator from Ref. \cite{TW}, which is essentially a combination of two filters, one forward-time and another reverse-time.

\subsection{Forward Filter}

The forward filter has the following form \cite{RPH}:
\begin{equation}
\Theta_{-}(t) = \chi \int_{-\infty}^t \! \theta(\zeta) e^{\chi.(\zeta-t)} d\zeta = \chi \left( \theta(t) * e^{-\chi .t} \right),
\end{equation}
where the value of $\chi$ is $\chi_{opt} = 2|\alpha|\sqrt{\kappa}$, which is optimal in the limit $\lambda \to 0$. Also, $\theta(t)$ comprise the measurements given by:
\begin{equation}\label{eq:measurement}
\theta(t) = \phi(t) + \frac{1}{2|\alpha|}w(t),
\end{equation}
where $w(t)$ is a zero-mean white Gaussian noise with unity amplitude.

The transfer function of this filter is:
\begin{equation}
\frac{\Theta_{-}(s)}{\theta(s)} = \frac{\chi}{s + \chi}.
\end{equation}

Thus, the forward-time process and filter equations are:

\begin{align}
\textsf{Process:}& \quad \dot{\phi} = -\lambda \phi + \sqrt{\kappa}v, \\ 
\textsf{Filter:}& \quad \dot{\Theta}_{-} = - \chi_{opt}\Theta_{-} + \chi_{opt}\phi + \frac{\chi_{opt}}{2|\alpha|}w.
\end{align}

Let the steady-state state covariance matrix for the forward system be:
\begin{equation}
\mathbf{P_{fs}} = \left[ \begin{array}{cc}
\Sigma & M_f \\
M_f & N_f
\end{array} \right].
\end{equation}

Note that
\[ \mathbf{P_{fs}} = E\left\lbrace
\left[ \begin{array}{c}
\phi \\
\Theta_{-}
\end{array} \right]
\left[ \begin{array}{cc}
\phi & \Theta_{-}
\end{array} \right] \right\rbrace = E\left[ \begin{array}{cc}
\phi^2 & \phi\Theta_{-} \\ 
\Theta_{-}\phi & \Theta_{-}^2
\end{array} \right].
\]

Thus, we have
\[ \Sigma = E[\phi^2], \quad M_f = E[\phi\Theta_{-}], \quad N_f = E[\Theta_{-}^2]. \]

Upon solving the Lyapunov equation of the form (\ref{eq:lyapunov}) for the forward system, we get:
\begin{align}
\Sigma & = \frac{\kappa}{2\lambda},\\
M_f & = \frac{\chi_{opt}\kappa}{2\lambda(\lambda + \chi_{opt})},\\
N_f & = \frac{4\chi_{opt}|\alpha|^2\kappa + \chi_{opt}\lambda^2 + \chi_{opt}^2\lambda}{8|\alpha|^2\lambda(\lambda + \chi_{opt})}.
\end{align}

Thus, we obtain:
\begin{equation}\label{eq:fwd_err_cov_prl}
\boxed{\sigma_f^2 = \frac{\chi_{opt}(\lambda + 2\chi_{opt})}{8|\alpha|^2(\lambda + \chi_{opt})} = \frac{\sqrt{\kappa}(\lambda + 4|\alpha|\sqrt{\kappa})}{4|\alpha|(\lambda + 2|\alpha|\sqrt{\kappa})}.}
\end{equation}

One can verify that this expression for the error covariance agrees with $\sigma_{-}^2$ of Eq. (10) from Ref. \cite{TW} for the optimal case of $\chi$.

\subsection{Backward Filter}\label{sec:bwd_filter_prl}

The backward filter has the following form:
\begin{equation}
\Theta_{+}(t) = \chi \int_t^{\infty} \! \theta(\zeta) e^{-\chi.(\zeta-t)} d\zeta.
\end{equation}

Let $\zeta = T-\tilde{\zeta}$ and $\tau = T-t$. Then, $d\zeta = -d\tilde{\zeta}$. When $\zeta = t$, $\tilde{\zeta} = T-t = \tau$ and when $\zeta = \infty$, $\tilde{\zeta} = T - \infty = -\infty$.

Thus, we get:
\begin{align*}
\Theta_{+}(\tau) &= - \chi \int_{\tau}^{-\infty} \! \theta(T-\tilde{\zeta}) e^{-\chi.(\tau - \tilde{\zeta})} d\tilde{\zeta} \\ 
\Rightarrow \Theta_{+}(\tau) &= \chi \int_{-\infty}^{\tau} \! \tilde{\theta}(\tilde{\zeta}) e^{-\chi.(\tau - \tilde{\zeta})} d\tilde{\zeta} = \chi \left( \tilde{\theta}(\tau) * e^{-\chi.\tau} \right),
\end{align*}
where $\tilde{\theta}(\tilde{\zeta}) = \theta(T - \tilde{\zeta})$.

Thus, we get in the Laplace domain:
\begin{equation}
\frac{\Theta_{+}(s)}{\tilde{\theta}(s)} = \frac{\chi}{s + \chi},
\end{equation}
where $s$ is the Laplace variable.

Thus, we obtain:

\[ \dot{\Theta}_{+}(\tau) = -\chi \Theta_{+}(\tau) + \chi \tilde{\theta}(\tau), \]
where again the value of $\chi$ is $\chi_{opt} = 2|\alpha|\sqrt{\kappa}$, which is optimal in the limit $\lambda \to 0$.

When our model (\ref{eq:ou_noise}),(\ref{eq:measurement}), which is driven by Gaussian white noise, has reached steady state, the output process will be a stationary Gaussian random process, which is described purely by its auto-correlation function. If we consider this output process in reverse time, this will also be a stationary random process with the same auto-correlation function. This follows from the definition of the auto-correlation function. Hence, the statistics of the reversed time output process are the same as the statistics of the forward time output process. Thus, the reversed time output process can be regarded as being generated by the same (and not time reversed) process that generated the forward time process, i.e.

\begin{align*}
\dot{\phi}(\tau) &= -\lambda\phi(\tau) + \sqrt{\kappa}v(\tau), \\ 
\tilde{\theta}(\tau) &= \phi(\tau) + \frac{1}{2|\alpha|}w(\tau).
\end{align*}

Now, we would use the Lyapunov method to deduce the state covariance matrix and error-covariance of the backward filter, the process and filter equations of which are:

\begin{align}
\textsf{Process:}& \quad \dot{\phi} = -\lambda \phi + \sqrt{\kappa}v, \\ 
\textsf{Filter:}& \quad \dot{\Theta}_{+} = -\chi_{opt}\Theta_{+} + \chi_{opt}\phi + \frac{\chi_{opt}}{2|\alpha|}w.
\end{align}

Let the steady-state state covariance matrix for the backward system be:
\begin{equation}
\mathbf{P_{bs}} = \left[ \begin{array}{cc}
\Sigma & M_b \\
M_b & N_b
\end{array} \right],
\end{equation}
where we have
\[ \Sigma = E[\phi^2], \quad M_b = E[\phi\Theta_{+}], \quad N_b = E[\Theta_{+}^2]. \]

Upon solving the Lyapunov equation of the form (\ref{eq:lyapunov}) for the information filter, we get:
\begin{align}
\Sigma & = \frac{\kappa}{2\lambda},\\
M_b & = \frac{\chi_{opt}\kappa}{2\lambda(\lambda + \chi_{opt})},\\
N_b & = \frac{4\chi_{opt}|\alpha|^2\kappa + \chi_{opt}\lambda^2 + \chi_{opt}^2\lambda}{8|\alpha|^2\lambda(\lambda + \chi_{opt})}.
\end{align}

Thus, we obtain:
\begin{equation}\label{eq:bwd_err_cov_prl}
\boxed{\sigma_b^2 = \frac{\chi_{opt}(\lambda + 2\chi_{opt})}{8|\alpha|^2(\lambda + \chi_{opt})} = \frac{\sqrt{\kappa}(\lambda + 4|\alpha|\sqrt{\kappa})}{4|\alpha|(\lambda + 2|\alpha|\sqrt{\kappa})}.}
\end{equation}

One can verify that this expression for the error covariance agrees with $\sigma_{+}^2$ of Eq. (10) from Ref. \cite{TW} for the optimal case of $\chi$.

\subsection{Smoothed Error Covariance}\label{sec:smoothed_err_cov}

Suppose we have two unbiased estimates of some state $x$. We call these $\hat{x}_1$ and $\hat{x}_2$. We form a new estimate $\hat{x}$ as a linear combination of $\hat{x}_1$ and $\hat{x}_2$ \cite{RGB}:
\begin{equation}
\hat{x} = k_1\hat{x}_1 + k_2\hat{x}_2,
\end{equation}
where for the new estimate to be unbiased,
\[ k_1 + k_2 = 1. \]

The mean-square error for $\hat{x}$ is then

\begin{equation}
E[(x-\hat{x})^2] = E\left\lbrace[x-k_1\hat{x}_1-(1-k_1)\hat{x}_2]^2\right\rbrace,
\end{equation}
or,
\small
\begin{equation}\label{eq:unbiased_error}
\begin{split}
E[e^2] &= E\left\lbrace[k_1(e_1-e_2)+e_2]^2\right\rbrace \\ 
&= k_1^2E[e_1^2]+(1-k_1)^2E[e_2^2]+2k_1(1-k_1)E[e_1e_2],
\end{split}
\end{equation}
\normalsize
where $e$, $e_1$, and $e_2$ are the errors in $\hat{x}$, $\hat{x}_1$, and $\hat{x}_2$, respectively. Eq. (\ref{eq:unbiased_error}) may now be differentiated with respect to $k_1$ and set equal to $0$ to find the optimal $k_1$. Thus, we obtain:

\begin{equation}
k_1 = \frac{E[e_2^2]-E[e_1e_2]}{E[e_1^2]+E[e_2^2]-2E[e_1e_2]}.
\end{equation}

Substituting for $k_1$ in (\ref{eq:unbiased_error}), we get:
\begin{equation}\label{eq:smoothed_error}
\boxed{E[e^2] = \frac{E[e_1^2]E[e_2^2]-(E[e_1e_2])^2}{E[e_1^2]+E[e_2^2]-2E[e_1e_2]}.}
\end{equation}

This relation can be used to obtain the smoothed error covariance, given the forward and backward systems. In our case, $e_1=\phi-\Theta_{-}$, $e_2=\phi-\Theta_{+}$, $E[e_1^2]=\sigma_f^2$, $E[e_2^2]=\sigma_b^2$ and it remains to evaluate $E[e_1e_2]=E[(\phi-\Theta_{-})(\phi-\Theta_{+})]$.

\begin{align*}
E[e_1e_2] &= E\left\lbrace \left[ \begin{array}{cc}
1 & -1
\end{array} \right] \left[ \begin{array}{c}
\phi \\
\Theta_{-}
\end{array} \right] \left[ \begin{array}{cc}
\phi & \Theta_{+}
\end{array} \right] \left[ \begin{array}{c}
1 \\
-1
\end{array} \right] \right\rbrace \\ 
&= \left[ \begin{array}{cc}
1 & -1
\end{array} \right] E \left[ \begin{array}{cc}
\phi^2 & \phi\Theta_{+} \\ 
\Theta_{-}\phi & \Theta_{-}\Theta_{+}
\end{array} \right] \left[ \begin{array}{c}
1 \\
-1
\end{array} \right] \\ 
&= \left[ \begin{array}{cc}
1 & -1
\end{array} \right] \left[ \begin{array}{cc}
\Sigma & M_b \\ 
M_f & \alpha\Sigma\beta
\end{array} \right] \left[ \begin{array}{c}
1 \\
-1
\end{array} \right] \\ 
&= \Sigma - M_f - M_b + \alpha\Sigma\beta ,
\end{align*}
where $\alpha = M_f\Sigma^{-1}$ and $\beta = \Sigma^{-1}M_b$ \cite{WWS}. Thus, we get:

\begin{equation}
\boxed{\sigma_{fb}^2 = E[(\phi-\Theta_{-})(\phi-\Theta_{+})] = \frac{\kappa\lambda}{2(\lambda+\chi_{opt})^2}.}
\end{equation}

Note that this agrees with Eq. (11) from Ref. \cite{TW} for the optimal case of $\chi$.

Upon substituting appropriately in (\ref{eq:smoothed_error}), we thus get the following as the smoothed error covariance:
\begin{equation}
\boxed{\sigma_s^2 = \frac{\sqrt{\kappa}(\lambda^2 + 8|\alpha|\lambda\sqrt{\kappa} + 8|\alpha|^2\kappa)}{8|\alpha| (\lambda + 2|\alpha|\sqrt{\kappa})^2}.}
\end{equation}

One can verify that this expression for the error covariance agrees with $\sigma^2$ of Eq. (12) from Ref. \cite{TW} for the optimal case of $\chi$.

\section{ADAPTIVE PHASE ESTIMATION USING A RAUCH-TUNG-STRIEBEL SMOOTHER}

In this section we design the optimal RTS smoother for the adaptive system of Ref. \cite{TW} and analyse the error-covariance of the same to show that it is equivalent to an optimal two-filter smoother, before we compare it with the estimator used in Ref. \cite{TW} and the Kalman filter used in Ref. \cite{RPH} in the next section.

\subsection{Forward Filter}

The forward filter is the same as the Kalman filter from Ref. \cite{RPH}. The steady-state Riccati equation is:

\begin{equation}
\boxed{-2\lambda P_f - 4|\alpha|^2 P_f^2 + \kappa = 0.}
\end{equation}

The stabilising solution of the above equation for $P_f$ is:
\begin{equation}\label{eq:fwd_err_cov_kalman}
\boxed{P_f = \frac{1}{4|\alpha|^2} \left( -\lambda + \sqrt{4\kappa|\alpha|^2 + \lambda^2} \right).}
\end{equation}

The Kalman gain is:
\begin{equation}
K_f = -\lambda + \sqrt{4\kappa|\alpha|^2 + \lambda^2}.
\end{equation}

The forward filter equation is:
\begin{equation}\label{eq:kalman_fwd_filter}
\boxed{ \dot{\hat{\phi}}_f = -(\lambda + K_f)\hat{\phi}_f + K_f\phi + \frac{K_f}{2|\alpha|}w.}
\end{equation}

\subsection{Rauch-Tung-Striebel Smoother}

The steady-state Riccati equation for the RTS smoother is \cite{LXP}:

\begin{equation}
\boxed{ -2\lambda P + 2\kappa P_f^{-1}P - \kappa = 0. }
\end{equation}

Upon substituting for $P_f$ and solving for $P$, we get:
\begin{equation}\label{eq:smoother_cov}
\boxed{P = \frac{\kappa}{2\sqrt{4\kappa |\alpha|^2 + \lambda^2}}.}
\end{equation}

Note that when $\lambda = 0$, we get $P = \sqrt{\kappa}/4|\alpha|$, as desired (see Ref. \cite{TW}).

The smoother gain is obtained to be \cite{LXP}:
\begin{equation}
F = \frac{4|\alpha|^2\kappa}{-\lambda + \sqrt{4\kappa |\alpha|^2 + \lambda^2}}.
\end{equation}

The equation for the smoothed estimate is \cite{LXP}:
\begin{equation}
\boxed{ \dot{\hat{\phi}} = (-\lambda + F)\hat{\phi} - F\hat{\phi}_f. }
\end{equation}

\subsection{Backward Filter}

The RTS smoother, as obtained in the previous section, abstracts away the backward filter, the error-covariance and the filter equation of which are explicitly derived here for reference later.

The steady-state Riccati equation for the backward filter is \cite{LXP}:
\begin{equation}
\boxed{2\lambda P_b - 4|\alpha|^2 P_b^2 + \kappa = 0.}
\end{equation}

The stabilising solution of the above equation for $P_b$ is:
\begin{equation}\label{eq:bwd_err_cov_kalman}
\boxed{P_b = \frac{1}{4|\alpha|^2} \left( \lambda + \sqrt{4\kappa|\alpha|^2 + \lambda^2} \right).}
\end{equation}

Thus, the Kalman gain for the backward filter is:
\begin{equation}
K_b = \lambda + \sqrt{4\kappa|\alpha|^2 + \lambda^2}.
\end{equation}

The backward filter equation is:
\begin{equation}\label{eq:kalman_bwd_filter}
\boxed{ \dot{\hat{\phi}}_b = (\lambda - K_b)\hat{\phi}_b + K_b\phi + \frac{K_b}{2|\alpha|}w.}
\end{equation}

\paragraph*{Remark.} One can verify that the error-covariances of the forward and the backward Kalman filters, obtained using the Lyapunov method used in section \ref{sec:ape_prl}, would be the same as in (\ref{eq:fwd_err_cov_kalman}) and (\ref{eq:bwd_err_cov_kalman}), respectively. In addition, referring to section \ref{sec:smoothed_err_cov}, one can show:
\begin{equation}
E[(\phi-\hat{\phi}_f)(\phi-\hat{\phi}_b)] = 0,
\end{equation}
which implies that the forward and the backward estimates of the optimal smoother are independent. The error covariance of the RTS smoother thus obtained from (\ref{eq:smoothed_error}) would agree with (\ref{eq:smoother_cov}).

\section{COMPARISON BETWEEN RTS SMOOTHER, KALMAN FILTER USED IN REF. \cite{RPH}, AND FILTER AND SMOOTHER USED IN REF. \cite{TW}}

\begin{figure}[!t]
\hspace*{-8mm}
\includegraphics[width=0.57\textwidth]{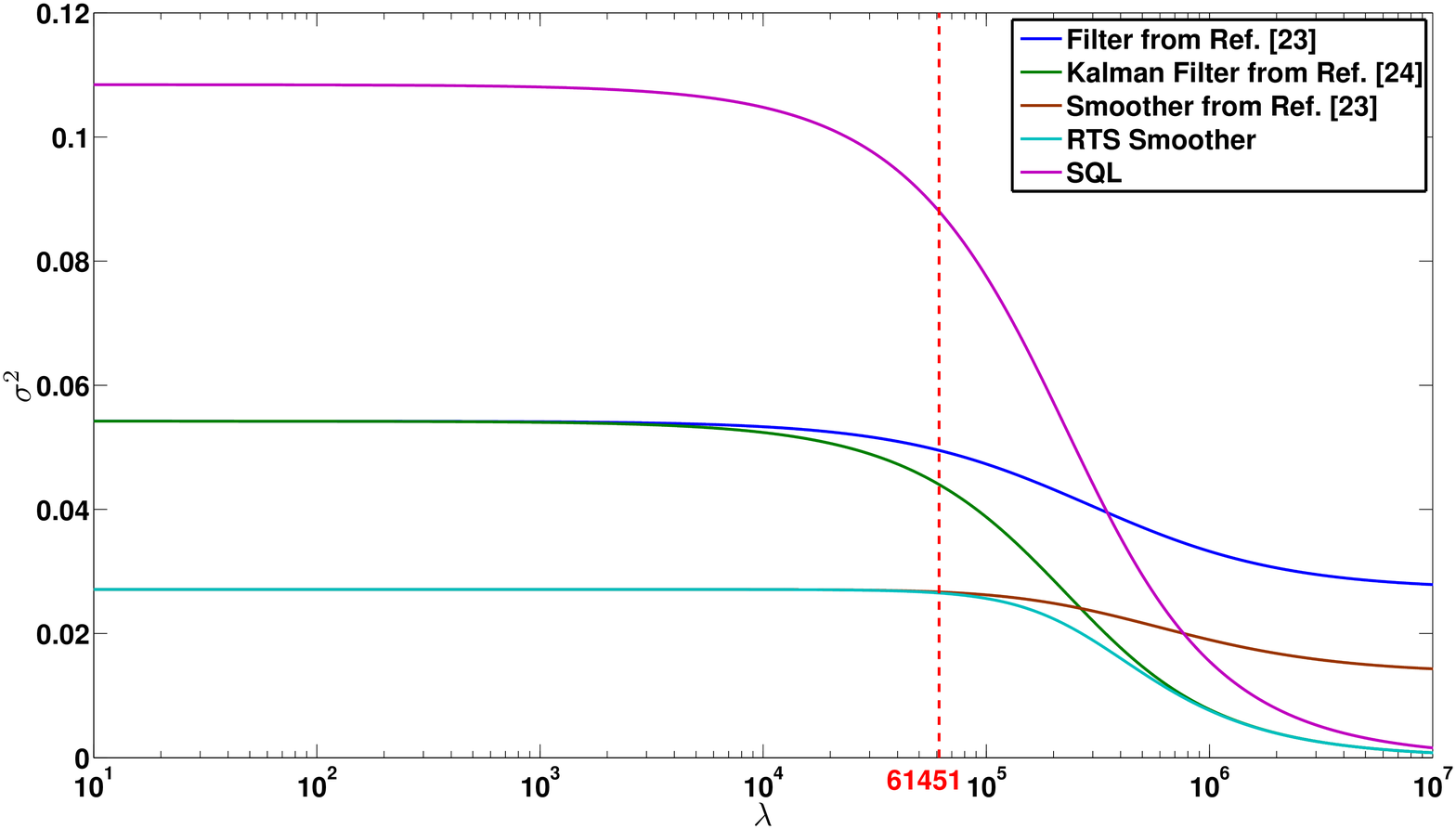}
\caption{Comparison of the error covariance between the RTS smoother, Kalman filter used in Ref. \cite{RPH}, and filter and smoother used in Ref. \cite{TW}.}
\label{fig:comparison_graph}
\end{figure}

Fig. \ref{fig:comparison_graph} shows the plot of the mean-square error against the parameter $\lambda$ for the four cases, viz. RTS smoother, Kalman filter used in Ref. \cite{RPH}, filter used in Ref. \cite{TW} and smoother used in \cite{TW}, as compared to the SQL for the noise process modulating the signal phase being estimated. The nominal experimental values used in the adaptive experiment in Ref. \cite{TW} were used for the other parameters in obtaining these graphs. It is clear that smoothing offers improvement over filtering alone in the accuracy of the estimate. For lower values of $\lambda$, both the filter and the smoother from Ref. \cite{TW} approximate well the Kalman filter and RTS smoother, respectively. However, with increasing $\lambda$, the optimal filter and smoother improve significantly and beat the SQL throughout, unlike the sub-optimal ones used in Ref. \cite{TW}. In summary, it is evident that the RTS smoother is the best scheme. The red vertical line denotes the value of $\lambda$ used in the adaptive experiment in Ref. \cite{TW}.

\section{ADAPTIVE PHASE ESTIMATION USING A ROBUST FIXED-INTERVAL SMOOTHER}

We shall use the technique laid down in Ref. \cite{MSP} to build a robust smoother for our continuous-time uncertain system, satisfying a certain Integral Quadratic Constraint as in Eq. (2.4) in Ref. \cite{MSP}. The uncertainty is introduced in the parameter $\lambda$ as follows:
\[ \lambda \to \lambda - \mu\Delta\lambda, \]
where $\Delta$ is an uncertain parameter satisfying $|\Delta| \leq 1$, and $0 \leq \mu < 1$ determines the level of uncertainty in the model.

Eq. (2.5) in Ref. \cite{MSP}, then, takes the form:
\begin{align}
\textsf{Process:}& \quad \dot{\phi} = -\lambda\phi + B_1\Delta K \phi + B_1 v, \\ 
\textsf{Measurement:}& \quad \theta = \phi + \frac{1}{2|\alpha|}w,
\end{align}
where $B_1 = \sqrt{\kappa}$ and $K = \mu\lambda/\sqrt{\kappa}$.

The \emph{uncertainty output} of Eq. (2.1) in Ref. \cite{MSP} for our system is:
\begin{equation}
z = \frac{\mu\lambda}{\sqrt{\kappa}}\phi.
\end{equation}

For the purpose of the Integral Quadratic Constraint (IQC) satisfied by our system, we will have $X_0 = 0$, since no \emph{a-priori} information exists about the initial condition of the state in our case. Also, $Q=1$ for the uncertainty matrix $\Delta$ to satisfy the required bound. Also, $d=1$, since the amplitudes of the white noise processes $v$ and $w$ have been assumed to be unity. The IQC of Eq. (2.4) in Ref. \cite{MSP}, thus, takes the following form in our case:

\begin{equation}
\int_0^T \! (\tilde{w}^2 + \frac{1}{4|\alpha|^2}\tilde{v}^2R) dt \leq 1 + \int_0^T \! ||z||^2 dt,
\end{equation}
where $\tilde{w}=\Delta K\phi+v$ and $\tilde{v}=w$ are the {\it uncertainty inputs}.

Thus, we would have $R = 4|\alpha|^2$ in our case.

The steady-state forward Riccati equation, as obtained from Eq. (5.1) in Ref. \cite{MSP} for our case, is:
\begin{equation}\label{eq:robust_fwd_riccati}
-2\lambda X + \kappa X^2 + \frac{\mu^2\lambda^2}{\kappa} - 4|\alpha|^2 = 0.
\end{equation}

The stabilising solution of the above equation for $X$ is:
\begin{equation}
X = \frac{\lambda + \sqrt{\lambda^2 - \mu^2\lambda^2 + 4|\alpha|^2\kappa}}{\kappa}.
\end{equation}

The steady-state backward Riccati equation, as obtained from Eq. (5.2) in Ref. \cite{MSP} for our case, is:
\begin{equation}\label{eq:robust_bwd_riccati}
-2\lambda Y - \kappa Y^2 - \frac{\mu^2\lambda^2}{\kappa} + 4|\alpha|^2 = 0.
\end{equation}

The stabilising solution of the above equation for $Y$ is:
\begin{equation}
Y = \frac{-\lambda + \sqrt{\lambda^2 - \mu^2\lambda^2 + 4|\alpha|^2\kappa}}{\kappa}.
\end{equation}

Next, Eq. (5.3) in Ref. \cite{MSP} for our case yields:
\begin{equation}
\dot{\eta} = -(\sqrt{\lambda^2 - \mu^2\lambda^2 + 4|\alpha|^2\kappa})\eta + 4|\alpha|^2\phi + 2|\alpha|w.
\end{equation}

Likewise, Eq. (5.4) in Ref. \cite{MSP} for reverse-time yields:
\begin{equation}
\dot{\xi} = -(\sqrt{\lambda^2 - \mu^2\lambda^2 + 4|\alpha|^2\kappa})\xi + 4|\alpha|^2\phi + 2|\alpha|w.
\end{equation}

The forward filter is, then, simply the centre of the ellipse of Eq. (3.3) in Ref. \cite{MSP}:
\begin{equation}
\hat{\phi}_f = \frac{\eta}{X}.
\end{equation}

Likewise, the backward filter is:
\begin{equation}
\hat{\phi}_b = \frac{\xi}{Y}.
\end{equation}

The forward filter differential equation is, thus:
\begin{equation}\label{eq:robust_fwd_filter}
\boxed{\dot{\hat{\phi}}_f = -L\hat{\phi}_f + \frac{4|\alpha|^2\kappa}{\lambda + L}\phi + \frac{2|\alpha|\kappa}{\lambda + L}w,}
\end{equation}
and the backward filter differential equation is:
\begin{equation}\label{eq:robust_bwd_filter}
\boxed{\dot{\hat{\phi}}_b = -L\hat{\phi}_b + \frac{4|\alpha|^2\kappa}{-\lambda + L}\phi + \frac{2|\alpha|\kappa}{-\lambda + L}w,}
\end{equation}
where $L = \sqrt{\lambda^2 - \mu^2\lambda^2 + 4|\alpha|^2\kappa}$.

The robust smoother for the uncertain system would, then, be the centre of the ellipse of Eq. (5.5) in Ref. \cite{MSP}:
\begin{equation}
\hat{\phi} = \frac{\eta - \xi}{X + Y}.
\end{equation}

One can verify that for $\mu=0$, (\ref{eq:robust_fwd_filter}) and (\ref{eq:robust_bwd_filter}) reduce to (\ref{eq:kalman_fwd_filter}) and (\ref{eq:kalman_bwd_filter}), respectively.

\section{COMPARISON BETWEEN ROBUST AND RTS SMOOTHERS FOR THE UNCERTAIN SYSTEM}

\begin{figure}[!b]
\hspace*{-5mm}
\includegraphics[width=0.56\textwidth]{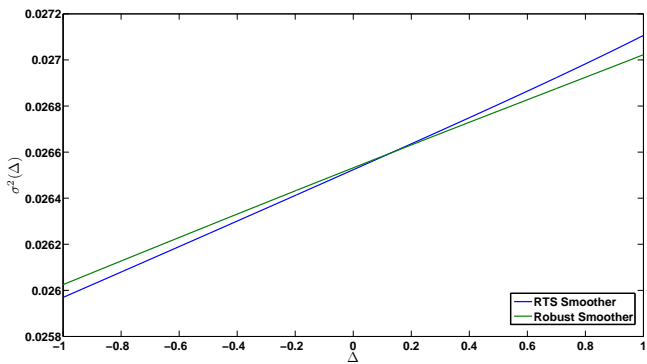}
\caption{Comparison of error covariance as a function of $\Delta$ for $\mu = 0.5$.}
\label{fig:robust_vs_rts1}
\end{figure}

The error-covariances of the robust smoother and the RTS smoother for the uncertain system may be computed using the Lyapunov technique employed in section \ref{sec:ape_prl}, as a function of $\Delta$, for the nominal experimental values of all the parameters and a given value of $\mu$. Fig. \ref{fig:robust_vs_rts1} shows the comparison of the two for the case of $50\%$ uncertainty, below which the robust smoother is not quite significantly superior in performance as compared to the RTS smoother.

\begin{figure}[!htb]
\hspace*{-5mm}
\includegraphics[width=0.56\textwidth]{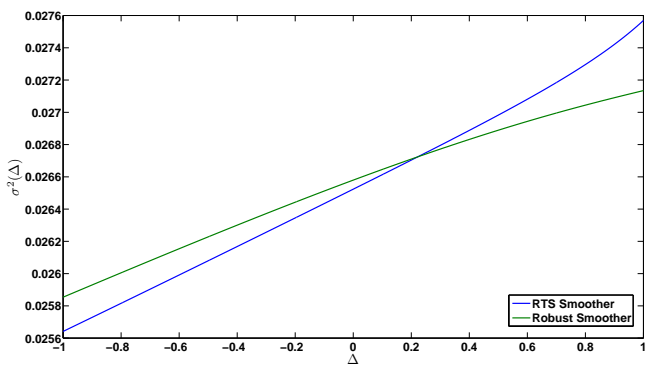}
\caption{Comparison of error covariance as a function of $\Delta$ for $\mu = 0.8$.}
\label{fig:robust_vs_rts2}
\end{figure}

\begin{figure}[!htb]
\hspace*{-5mm}
\includegraphics[width=0.56\textwidth]{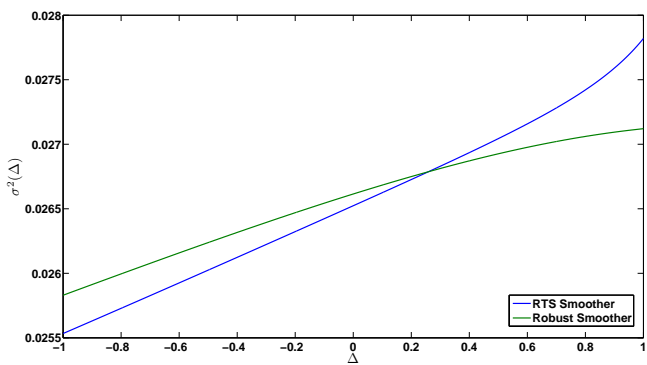}
\caption{Comparison of error covariance as a function of $\Delta$ for $\mu = 0.9$.}
\label{fig:robust_vs_rts3}
\end{figure}

Figs. \ref{fig:robust_vs_rts2} and \ref{fig:robust_vs_rts3} show the comparison for $\mu = 0.8$ and $\mu = 0.9$, respectively. Clearly, the robust smoother performs much better than the RTS smoother as $\Delta$ approaches $1$ for all levels of uncertainty in $\lambda$.

\section{CONCLUSION}
This paper extends the optimal and robust filtering theory of Ref. \cite{RPH}, as applied to adaptive continuous homodyne phase estimation of a coherent state of light, to include optimal RTS and robust fixed-interval smoothing rather than filtering alone. In particular, it presents an insightful analysis of the relative performance of these various schemes with respect to the standard quantum limit. These theoretical results are to be demonstrated experimentally as part of further work. It would be interesting to extend these results for the case of squeezed states of light or other complex noise processes. Robustness to uncertainties in other parameters such as the photon flux or the noise power may also be explored.

\bibliographystyle{IEEEtran}
\bibliography{IEEEabrv,robustbiblio}

\end{document}